\documentclass[aps,superscriptaddress,showpacs,prl,nofootinbib,twocolumn]{revtex4-1}

\usepackage{graphics} 
  
\usepackage{xcolor}
\usepackage{epsfig} 
\input{epsf}
\usepackage{natbib} 
\usepackage[colorlinks=true,citecolor=blue,linkcolor=blue,urlcolor=blue]{
hyperref}
\setcitestyle{square, numbers, comma, sort&compress}

\def\be{\begin{equation}}
\def\ee{\end{equation}}
\def\bea{\begin{eqnarray}}
\def\eea{\end{eqnarray}}
\def\ms{\overline {\rm MS}}

\begin{document} 
 
\title{QCD pressure: renormalization group optimized perturbation theory confronts lattice}

\author{Jean-Lo\"{\i}c Kneur}\email{jean-loic.kneur@umontpellier.fr}
\affiliation{Laboratoire Charles Coulomb (L2C), UMR 5221 CNRS-Universit\'{e} Montpellier, 34095 Montpellier, France}  
\author{Marcus Benghi Pinto} \email{marcus.benghi@ufsc.br}
\affiliation{Departamento de F\'{\i}sica, Universidade Federal de Santa
  Catarina, 88040-900 Florian\'{o}polis, SC, Brazil}
 \author{Tulio E. Restrepo} \email{tulio.restrepo@posgrad.ufsc.br}
\affiliation{Departamento de F\'{\i}sica, Universidade Federal de Santa
  Catarina, 88040-900 Florian\'{o}polis, SC, Brazil}
 
 \begin{abstract} 
 The quark contribution to the QCD pressure, $P_q$, is evaluated up to next-to-leading order (NLO)
 within the renormalization group optimized perturbation theory (RGOPT) resummation approach. 
 To evaluate the complete QCD pressure we simply add the perturbative NLO contribution from massless 
 gluons to the resummed $P_q$. Despite of this unsophisticated approximation our results for 
 $P = P_q +P_g$ at the central scale $M\sim 2\pi T$ show a remarkable agreement with lattice 
 predictions for $0.25 \lesssim T \lesssim 1 \, {\rm GeV}$. We also show that by being imbued with RG properties, the RGOPT 
 produces a drastic reduction of the embarrassing remnant scale dependence that plagues both standard 
 thermal perturbative QCD and hard thermal loop perturbation theory (HTLpt) applications.
 \end{abstract} 
 
 
 \maketitle
 At the fundamental level, strongly interacting matter composed by quarks and gluons is described by the 
 non Abelian theory  of quantum chromodynamics  (QCD) whose coupling constant, $\alpha_s$, is predicted to decrease 
 with increasing energies as the system evolves to a regime of asymptotic freedom (AF). This property, together with 
 the other crucial phenomena of confinement and chiral symmetry, has triggered the possibility of studying eventual phase 
 transitions related to (de)confinement and chiral symmetry breaking/restoration in the laboratory through 
 experiments involving heavy ion collisions.
 On the theoretical side, lattice QCD (LQCD) ab initio simulations have predicted that the deconfinement 
 and chiral symmetry restoration occur via an analytic crossover 
 occuring at a pseudo-critical temperature of order $T_{pc} \simeq 155 \, {\rm MeV}$ with the baryon 
 chemical potential, $\mu_B$, approaching zero \cite{lattice}. 
 The region at intermediate $T$ and $\mu_B$ values ($T \sim 100 \, {\rm MeV}$, $\mu_B \sim 900\, {\rm MeV} $), 
 is currently being explored by experiments such as the beam energy scan at RHIC whose aim is to confirm the existence 
 of a critical end point which locates the end of a first order transition line predicted to start at $T=0$ \cite{buballa}.
 Another region, covering 
 the range $T \sim 0-30 \,{\rm MeV}$ and $\mu_B \gtrsim 1000\, {\rm MeV}$, is essential for the description of 
 compact stellar objects such as neutron stars. Unfortunately, due to the notorious sign problem \cite{sign}, 
 LQCD  encounters a more hostile environment within these two phenomenologically important regions where numerical
 simulations cannot yet be reliably implemented. 
Therefore, the development of reliable alternatives with more analytical tools 
remains timely. 
 Several such alternatives can partly address the
deconfinement and/or chiral symmetry restoration, like extensions of the Nambu--Jona-Lasinio (NJL) model\cite{buballa,NJLCosta,PNJL,PNJLCosta}, 
or with more complete QCD dynamics,  
the Dyson-Schwinger equations (see e.g. \cite{DSE,DSErecent}), the functional Renormalization Group~\cite{fRG}, or other approaches\cite{others}.
Our present approach is rather built on weak-coupling expansion 
as a starting point in the evaluation of physical observables in powers (and logarithms) of $g=4\pi \alpha_s$. 
However, perturbative results require
a further resummation to be compatible with strong or even moderate coupling regimes (see Refs.
\cite{Trev,laine,HTLrev2020} for reviews). At finite temperatures, resumming the perturbative series 
cure some of the infrared divergences from zero modes, improving also convergence issues (but does not solve the intrinsically  
nonperturbative infrared issues due to static magnetic fields\cite{Linde}). 
An efficient way to perform a resummation is to reorganize the perturbative series around a quasiparticle mass parameter. 
Such an approach appears in the literature under various names, 
like optimized perturbation theory (OPT)~\cite{pms,opt_phi4,opt_qcd}, 
linear $\delta$ expansion (LDE)~\cite{lde}, variational perturbation theory (VPT)~\cite{vpt}, 
or in the thermal context, screened perturbation theory (SPT) \cite{spt,spt4L}. \\
Analoguous thermal resummations in the QCD gluon sector is far from obvious due to gauge-invariance issues, but 
had been circumvented by Braaten and Pisarski~\cite{HTLbasic} who proposed 
a gauge-invariant non-local Lagrangian embedding hard thermal loop (HTL)
contributions, Landau damping, and screening gluon thermal mass, with momentum-dependent self-energies and HTL-dressed vertices. 
The high temperature approximation of HTL could be successfully generalized in the so-called HTL
perturbation theory (HTLpt) \cite{HTLpt}, allowing for the evaluation of the QCD thermodynamics at the 
NNLO (three-loop), considering both the glue\cite{HTLptg3L} and quark sectors at finite temperatures and baryonic 
densities \cite {HTLptqcd2L,HTLptDense3L, HTLptqcd3L}. The final NNLO results turned out to be in good agreement 
with LQCD predictions for temperatures down to $T \approx 1.5 \, T_{pc}$ for the ``central" renormalization 
scale choice $M = 2\pi \sqrt  {T^2 + \mu^2/\pi^2}$. Unfortunately this  agreement  quickly deteriorates 
as moderate scale variations of a factor 2  induce relative variations of order 1 or more. 
Moreover it has been observed that this scale dependence  strongly {\it increases}  
at higher orders, most predominantly from NLO to NNLO. 
It is important to remark that standard pQCD results are  also plagued by a similar growing and strong scale 
dependence at high orders \cite{Trev,pQCD4L,pQCDmu4L}. 

More recently an alternative resummation approach
 has been proposed, the renormalization group optimized 
perturbation theory (RGOPT) \cite{JLGN,JLalphas,prlphi4,prdphi4}, that  essentially combines a variational mass prescription 
with embedded consistent RG invariance properties. 
Within QCD, at $T=\mu_B=0$, the method has been used to
estimate the coupling $\alpha_s$, predicting
values \cite{JLalphas}  compatible with the
world averages~\cite{PDG2018}. Still at $T=0=\mu$,  a precise
prediction was obtained for the quark condensate~\cite{JLcond,JLcond2}.
In thermal theories the RGOPT has been applied, e.g., 
to the simpler scalar $\phi^4$ model \cite{prlphi4,prdphi4} at NLO,
showing how it improves the generic residual scale dependence as compared to both 
standard thermal perturbation theory and SPT. 
Concerning QCD, the direct application of RGOPT to the pure glue sector is however momentarily obstructed by 
specific technical difficulties, as it involves new types of very involved thermal integrals \cite{gluons}.
Yet at least a nontrivial NLO evaluation of the QCD pressure 
can be performed in a more simple-minded and relatively easy way, if one considers the 
case of  massive quarks and massless gluons. 
In this vein we have recently applied the RGOPT to the QCD quark sector only, while considering
the gluons to be massless, in order to evaluate the NLO pressure at finite densities and vanishing
temperatures~\cite {prdCOLD}. Our NLO results show a good numerical agreement with {\em higher} order state of the art 
${\cal O}(g^3\ln^2 g)$ pQCD predictions, with a visible improvement (although relatively modest for cold matter) 
of the residual scale dependence. 
In the present work, we aim to extend the $T=0,\mu_B\ne0$  application performed in Ref. \cite {prdCOLD} in order to consider
a thermal bath. 
Our strategy is to apply our construction 
 to the quark sector which, together with the purely perturbative NLO contribution of massless gluons, 
will compose our complete NLO QCD pressure: $P(T,\mu_B)=P_q^{RGOPT}+P_g^{PT}$ 
where
%\footnote { Notice that since $P_g^{PT}$ vanishes at $T=0$ this is essentially the same  procedure adopted 
%in Ref. \cite {prdCOLD}. } 
$P_g^{PT} \sim T^4$. 
We believe that the results reported here represent a significant step  
towards the determination of thermodynamical observables  with RG improved properties. Technical details 
related to our calculation may be found in a companion paper~\cite {companion}.

Our starting point is the perturbative QCD pressure for three quark flavors with degenerate masses, $m_u=m_d=m_s\equiv m$,
and massless gluons: $P=P_q^{PT} + P_g^{PT}$. 
Let us consider first the quark contribution $P_q$ and how the RGOPT is built on it.
At NLO (${\cal O}(g)$) the {\em massive} quark  contributions to the pressure
can be obtained by combining the vacuum 
results of Ref. \cite {JLcond} and $T,\mu \ne 0$ results of Refs. \cite{kapusta-gale,laine2}. The {\it per flavor}
result reads
\begin{eqnarray}
\frac{P^{PT}_q}{N_f \,N_c}&=&-\frac{m^4}{8 \pi^2} \left(\frac{3}{4}
-L_m\right)+2T^4 J_1 \nonumber \\
&-&3g \frac{m^4}{2\left(2\pi\right)^4} C_F \left(L_m^2-\frac{4}{3}L_m+\frac{3}{4}\right)\nonumber \\
&-&g C_F \left \{ \left[\frac{m^2}{4\pi^2}  \left(2-3L_m \right) + \frac{T^2}{6}  
\right] T^2 J_2 \right . \nonumber\\
&+& \left . \frac{T^4}{2} J^2_2 +m^2 
T^2 J_3  \right \} \;,
\label{PPT}
\end{eqnarray}
where $L_m= \ln(m/M)$, $g\equiv 4 \pi \alpha_s(M)$, $M$ is the arbitrary 
renormalization scale in the $\ms$-scheme, $C_F = (N_c^2-1)/(2 N_c)$, $N_c=3$, and $N_f=3$.
In-medium and thermal effects are included in the (dimensionless) single integrals:
\begin{equation}
 J_1=\int\frac{ d^3{\bf \hat p}}{\left(2\pi\right)^3}
 \left\lbrace\ln\left[1+e^{-\left(E_p+\frac{\mu}{T}\right)}\right]+\ln\left[1+e^{-\left(E_p-\frac{\mu}{T}\right)}\right]\right\rbrace ,
\end{equation}
with ${\bf \hat p}\equiv {\bf p}/T$, $E_p=\sqrt{{\bf \hat p}^2 +m^2/T^2}$,  
\begin{equation}
 J_2=\int\frac {d^3 {\bf \hat p}}{\left(2\pi\right)^3}\frac{1}{E_p}
 \left[f^{+}(E_p)+f^-(E_p)\right]\;,
\end{equation}
and in the double integral (after angular integration over $p\cdot q/(|p||q|)$)
\begin{eqnarray}
J_3&=&\frac{1}{(2\pi)^4} \int^\infty_0 \int^\infty_0 \frac{d\hat p\, \hat p d\hat q 
\,\hat q}{E_p E_q} 
\left\{  \Sigma_+ \ln\left[\frac{E_p E_q -\frac{m^2}{T^2}-\hat p \hat q}{E_p E_q -\frac{m^2}{T^2}+\hat p \hat q}\right]\right . \nonumber \\
&+& \left . \Sigma_- \ln\left[\frac{E_p E_q +\frac{m^2}{T^2}+\hat p \hat q}{E_p E_q +\frac{m^2}{T^2} -\hat p \hat q}\right]
\right \} \;,
\end{eqnarray}
where $\Sigma_{\pm}=f^+\left(E_p\right)f^{\pm}\left(E_q\right)+f^-\left(E_p\right)f^\mp\left(E_q\right)$.

The Fermi-Dirac distributions  for anti-quarks ($+$ sign) and quarks ($-$ sign) read
\begin{equation}
f^\pm(E_p)=\frac{1}{1+e^{(E_p\pm \frac{\mu}{T})}} \;,
\end{equation}
where $\mu$ is the quark chemical potential, related to the baryonic chemical potential via $\mu_B = 3 \mu$. 
In the present work we consider symmetric quark matter and thus do not distinguish the chemical potentials 
associated with different flavors ($\mu_s=\mu_u=\mu_d\equiv \mu$). 
For the quark sector the Stefan-Boltzmann limit is
\begin{equation}
\frac{P_q^{SB}}{N_c N_f}= T^4  \left ( \frac {7 \pi^2}{180} \right )
\left ( 1 + \frac{120}{7} {\hat \mu}^2 + \frac{240}{7} {\hat \mu}^4 \right ) \;,
\end{equation}
where $ {\hat \mu} = \mu / (2\pi T)$.

Defining the (massive) homogenous RG operator
\begin{equation}
M \frac{d}{d M}= M \,\partial_M + \beta(g)\,\partial_g  - m\,\gamma_m(g) \, \partial_m \;,
\label{RG}
\end{equation}
note that acting with Eq.(\ref{RG}) on the massive pressure Eq.(\ref{PPT}) leaves a
non RG-invariant term of {\em leading order},
$\sim m^4 \ln(M)$.
To restore a perturbatively RG-invariant massive pressure, 
we proceed as in Refs.\cite{JLcond,prlphi4,prdphi4} (or closer to the present case, as in Ref. \cite{prdCOLD}), 
subtracting a finite zero point contribution,
\begin{equation}
\frac{P^{RGPT}_q}{N_c N_f} = \frac{P^{PT}_q}{N_c N_f} - \frac{m^4}{g} \sum_k s_k g^k \; ,
\label{PRGPT}
\end{equation}
where the $s_i$ are determined at successive orders so that 
\begin{equation}
M \frac{d}{d M} \left (\frac{P^{RGPT}_q}{N_c N_f} \right )   ={\cal O}(g^2 m^4) ,
\label{RGeq}
\end{equation}
up to neglected higher order terms. 
Our evaluations being carried up to ${\cal O}(g)$ it suffices to determine the first two 
$s_0$ and $s_1$ coefficients. They involve, through Eq.(\ref{RG}), coefficients of the 
$\beta$ function and anomalous mass dimension, $\gamma_m$, relevant to the $T=\mu=0$ pressure. 
Our normalizations are 
$\beta\left(g\right)=-2b_0g^2-2b_1g^3+\mathcal{O}\left(g^4\right)$
 and $\gamma_m\left(g\right)=\gamma_0g+\gamma_1g^2+\mathcal{O}\left(g^3\right)$ where
 \begin{eqnarray}
 \left(4\pi\right)^2 b_0&=& 11-\frac{2}{3}N_f ,\\
\left(4\pi\right)^4 b_1&=& 102-\frac{38}{3}N_f ,\\
 \gamma_0&=&\frac{1}{2\pi^2} \;,\;\;
\left(4\pi\right)^4 \gamma_1= \frac{404}{3}-\frac{40}{9}N_f \;.
\label{RGcoeff}
\end{eqnarray}
One then finds \cite{JLcond,prdCOLD} 
\be
s_0=-\left[(4\pi)^2 (b_0-2\gamma_0)\right]^{-1} ,
\label{s0def}
\ee 
 \begin{equation}
 s_1=-\frac{1}{4}\left[\frac{b_1-2\gamma_1}{4(b_0-2\gamma_0)} -\frac{1}{12\pi^2}\right ]\,.
 \label{s1def}
\end{equation}
Implementing the RGOPT involves the following steps (see for more details e.g. \cite{JLalphas,prdCOLD,companion}): 
1) first one restores (perturbative) RG invariance of the massive pressure, giving
Eq.(\ref{PRGPT}) with Eqs.(\ref{s0def}),(\ref{s1def}) at NLO. 2) The resulting expression is variationally modified, 
according to the prescription \cite{JLalphas,prdCOLD} 
\begin{equation}
P^{RGPT}(m\to m(1-\delta)^a, g\to \delta g) \equiv P^{RGOPT},
 \label{Lint}
\end{equation}
acting thus in the present case on Eq.(\ref{PRGPT}).
3) Next one reexpands (\ref{Lint}) in powers of $\delta$, setting $\delta\to 1$ to recover the massless case. 
 We stress that $m$ is now an arbitrary variational mass parameter, to be fixed by a sensible prescription
 explicited below.
 %\footnote{In principle 
 %the method can be generalized to accommodate physical current quark masses.}. 
4) At this stage one also needs to fix the exponent $a$ introduced in Eq.(\ref{Lint}), whose role
is crucial for RG consistency in our framework.
Expanding to  LO, ${\cal O}(\delta^0)$, and requiring the 
resulting pressure to satisfy the {\it reduced} (massless) RG equation: 
\begin{equation}
 \left [ M \partial_M  +\beta(g)\partial_g \right ]P^{RGOPT}_q = 0\; ,
 \label{RGred}
\end{equation}
leads to $a= \gamma_0/(2 b_0)$~\cite {JLalphas,JLcond,prdCOLD}.
 At higher orders, we keep for simplicity the same prescription, which has extra 
advantages as explained below.
The resulting NLO RGOPT-modified pressure after steps 1)--4) reads
\begin{eqnarray}
\frac{P^{RGOPT}_q}{N_fN_c}&=&-\frac{m^4}{8 \pi^2} \left(\frac{3}{4}-L_m\right)+2T^4\,J_1\nonumber \\
&+&\frac{m^4}{\left(2\pi\right)^2 }\left(\frac{\gamma_0}{b_0}\right)\left(\frac{1}{2}-L_m\right)+m^2
\left(\frac{\gamma_0}{b_0}\right) T^2\,J_2 \nonumber \\
 &-&3g \frac{m^4}{2\left(2\pi\right)^4} C_F \left(L_m^2-\frac{4}{3}L_m+\frac{3}{4}\right)\nonumber \\
 &-& g C_F \left \{ \left[\frac{m^2}{4\pi^2}  \left(2-3L_m \right) + \frac{T^2}{6}  
\right]  T^2J_2 \right . \nonumber \\
&+&\left . \frac{T^4}{2} J^2_2 + m^2 T^2 \, J_3 \right \} \nonumber \\ 
 &+&\frac{m^4}{\left(4\pi\right)^2 b_0}\left\lbrace \frac{1}{g}\left (1-\frac{\gamma_0}{b_0} \right )\right . \nonumber \\
&+& \left . \left[\left(b_1-2\gamma_1\right)\pi^2 -\frac{\left(b_0-2\gamma_0\right)}{3}\right]\right\rbrace\,.
\label{P2LRGOPT}
\end{eqnarray}
At NLO, Eq.(\ref{RGred}) with $a=\gamma_0/(2b_0)$ is no longer exactly satisfied, thus giving an independent constraint.
Accordingly in contrast with OPT/SPT the remnant $m$ can be fixed in two different manners\cite{prlphi4,prdphi4,prdCOLD}. 
Either, from using a standard stationarity criterion\cite{pms}, 
the mass optimization prescription (MOP): 
\begin{equation}
\frac{\partial {P}^{RGOPT}_q}{\partial m} \Bigr |_{\overline m} \equiv 0 \;,
\label{mop}
\end{equation}
or alternatively from Eq.(\ref{RGred}). 
The coupling $g(M)$ is 
determined from standard PT two-loop running, with renormalization scale $M$ chosen as 
a multiple of $\pi T$ as usual.
 At NLO $P_q^{RGOPT}(\overline m(g))$ inevitably has a remnant scale dependence,
basically because the subtractions in Eq.(\ref{PRGPT}) solely guarantee RG invariance up to
remnant higher order terms. 
But it is a nontrivial consequence of our subsequent construction, mainly Eq.(\ref{Lint}), that this remnant 
dependence remains moderate, of order $m^4 g^2$ at NLO, as will be illustrated below.
However, regarding Eq.(\ref{P2LRGOPT}), 
both Eq.(\ref{mop}) and Eq.(\ref{RGred}) fail to give a real dressed mass
$\overline m(g,T,\mu)$ for a substantial part of the physically relevant $T,\mu$ range.
The occurrence of non-unique solutions at higher orders, some being complex, is a well-known burden with OPT/SPT approaches.
In contrast $a=\gamma_0/(2b_0)$ in Eq.(\ref{Lint}) guarantees 
that the only acceptable solutions are those matching \cite{JLalphas} the asymptotic
freedom (AF) behavior for $g\to 0$ at $T=0$, 
a compelling criterion that often selects a unique solution.
In addition the nonreal solution issue can be cured in an RG consistent manner
by performing a renormalization scheme change (RSC)\cite {JLalphas, prdCOLD, companion}. 
With this aim we define a RSC acting only on the variational mass in our framework,
\begin{equation}
m \to m^\prime ( 1+ B_2 g^2) \,,
\label{RSC}
\end{equation}
where a single $B_2$ parametrizes a NLO RSC from the original ${\overline{\text{MS}}}$-scheme. 
The latter induces 
an extra term $- 4g m^4 s_0 B_2 $ in Eq.(\ref{PPT}) (renaming afterwards $m^\prime \to m$ 
the variational mass to avoid excessive notation changes), entering thus 
the MOP Eq.(\ref{mop}) and RG Eq.(\ref{RGred}). 
Now since we aim to solve the latter for exact $m$ and $g$ dependence, Eq.(\ref{RSC}) 
modifies those purposefully, when now considered as constraints for the arbitrary mass $m$
after the (all order) 
modifications induced from Eq.(\ref{Lint}).
Accordingly $B_2$ may be considered an extra variational parameter, quite similarly to $m$, 
thus to be fixed by a sensible prescription.\\
Considering specifically the RG Eq.(\ref{RGred}), it can be conveniently
written as a quadratic form in $\ln (m^2/M^2)$,
\begin{equation}
 -\ln \frac{ m^2}{M^2} + B_{rg}
 \mp \frac{8\pi^2}{g} \sqrt{\frac{2}{3} D_{rg}} = 0 \; , 
 \label{RGexact}
\end{equation}
where explicitly
\begin{equation}
 B_{rg} = 
 -\frac{1}{b_0\,g} +\frac{172}{81} -\frac{64}{81} \left (\frac{4g}{9\pi^2}\right )\,\frac{1}
 {1+\frac{4g}{9\pi^2}}
 +8\pi^2 \frac{T^2}{m^2} J_2 ,
 \label{Brg}
\end{equation}
\begin{eqnarray}
D_{rg} &=&-
\frac{g}{27}\frac{(4 g + 81 \pi^2)}{ (4g + 9\pi^2)^2}  -g^2 \left (\frac{3}{7} B_2  +\frac{11}{384\pi^4}\right )  \nonumber \\
&+& g^2\,
 \frac{T^4}{m^4} J_2 \left (J_2 -\frac{1}{6} \right )  - g^2\,\frac{T^2}{m^2} J_3 .
 \label{Drg}
 \end{eqnarray}
Note that a quite similar quadratic form can be obtained for the MOP Eq.(\ref{mop}), with its specific
expressions $B_{rg}\to B_{mop}$, $D_{rg}\to D_{mop}$ \cite{companion}.
As above anticipated, in the original $\ms$-scheme 
($B_2=0$) one can have $D_{rg} <0$ (or similarly $D_{mop}<0$) in some physically relevant 
parameter ranges due to some negative contributions,
leading thus to nonreal NLO $\overline m(T,\mu)$ solutions. 
Accordingly to recover real solutions in a large range, while at the same time fulfilling the crucial AF-matching requirement,
the comprehensive analysis performed in Ref. \cite {companion} 
suggests the following prescriptions: 
The arbitrary RSC parameter $B_2$ 
is fixed by partly (respectively fully) cancelling $D_{mop}$ (respectively $D_{rg}$).
For the RG prescription explicitly
\begin{equation}
D_{rg}(B_2) = 0 \,
 \label{B2sq0}
 \end{equation}
fixes $B_2$ using Eq.(\ref{Drg}). It gives  a single  
real solution for $\overline m$, 
determined by the first two terms of Eq.(\ref{RGexact}),  
the latter being still an implicit equation in $m$ for $T,\mu\ne 0$
via $J_2$ entering Eq.(\ref{Brg}).
The prescription Eq.(\ref{B2sq0}) may be seen at first as a rather peculiar choice,
but there happens to be very few other possible prescriptions to recover a real RG solution. 
Note that the resulting $\overline m(B_2)$ still involves arbitrary higher order contributions, 
as well as nontrivial $T,\mu$ dependence via  $B_{rg}$ in Eq.(\ref{Brg}). 
A similar analysis holds for the MOP Eq.(\ref{mop}), but 
leading to a $\overline m$ having quite different properties (we refer to Ref.\cite{companion} for details). 
We stress that for the two prescriptions the resulting $\overline m(B_2)$ is an intermediate 
variational parameter without much physical meaning outside its use in the pressure.
As illustrated below both prescriptions give drastically reduced 
remnant scale dependence as compared to pQCD, but the best results are obtained from
the RG prescription Eqs.(\ref{RGexact}),(\ref{B2sq0}). 
This is not very surprising as the latter more directly embeds RG properties as compared to Eq.(\ref{mop}).
However for more complete and conservative coverage of the possible variants at NLO, 
we will illustrate both RG and MOP prescription results below.   

\begin{figure}[h!]
\centerline{ \epsfig{file=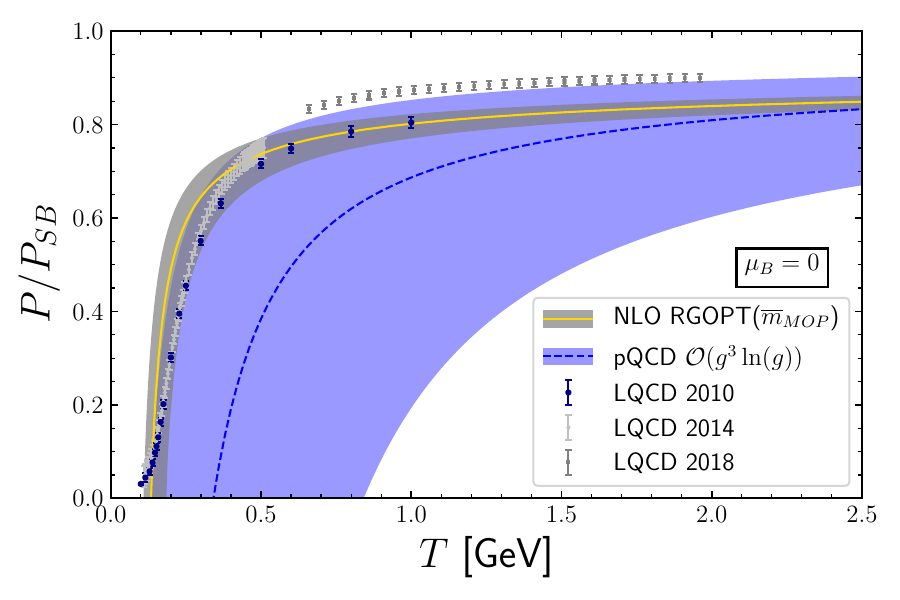,width=1.\linewidth,angle=0}}
\caption{NLO RGOPT (MOP prescription) plus NLO $P^{PT}_g$ pressure as function of $T$
at $\mu_B=0$ (grey band) 
compared to ${\rm N}^3LO\, g^3 \ln g$ pQCD (light blue band),
with scale dependence $\pi T \le M \le 4\pi T$,  and to lattice data \cite{LQCD2010,LQCD2014,LQCD2018}. }
\label{Pmu0bandg}
\end{figure}

Coming to the full QCD pressure, we simply add to Eq.(\ref{P2LRGOPT})
the NLO glue contributions \cite{Pglue},
\begin{equation}
P^{PT}_g= \frac{8\pi^2}{45} T^4 \, \left [ 1- \frac{15}{(4 \pi)^2} \, g \right ] \,.
\end{equation}
 Thus we can now compare the NLO RGOPT 
 full QCD results with those from HTLpt \cite{HTLptqcd2L,HTLptqcd3L} and (massless) pQCD \cite{pQCD4L}, 
 as well as with available LQCD data from Refs.~\cite{LQCD2010,LQCD2014,LQCD2018}. 
 For the numerical evaluations of NLO quantities we take the exact two-loop running coupling (see, e.g., Ref. \cite{prdCOLD}) 
 obtained by solving for $g(M)$
\begin{equation}
\ln \frac{M}{ \Lambda_{\overline{\text{MS}}} } = \frac{1}{2b_0\, g} +
\frac{b_1}{2b_0^2} \ln \left ( \frac{b_0 g} {1+\frac{b_1}{b_0} g} \right) ,
\label{g2L}
\end{equation}
for a given $\Lambda_{\overline{\text{MS}}}$ value (this also defines 
$\Lambda_{\overline{\text{MS}}}$ at two-loop level in our conventions).  
We take
$\Lambda_{\overline{\rm MS}}= 335 \,{\rm MeV}$ for $N_f=3$,
which is very close to the latest world average value\cite{PDG2018}.  
Notice that, for consistency  the NNLO HTLpt\cite {HTLptqcd3L} and ${\cal O}(g^3 \ln g)$ pQCD\cite{pQCD4L} 
numerical results reproduced here have been obtained rather with a three-loop order running coupling.\\
%%%%%%%%%%%%%%%%%%%%%%%%%%%%%%
\begin{figure}[h!]
    \centerline{ \epsfig{file=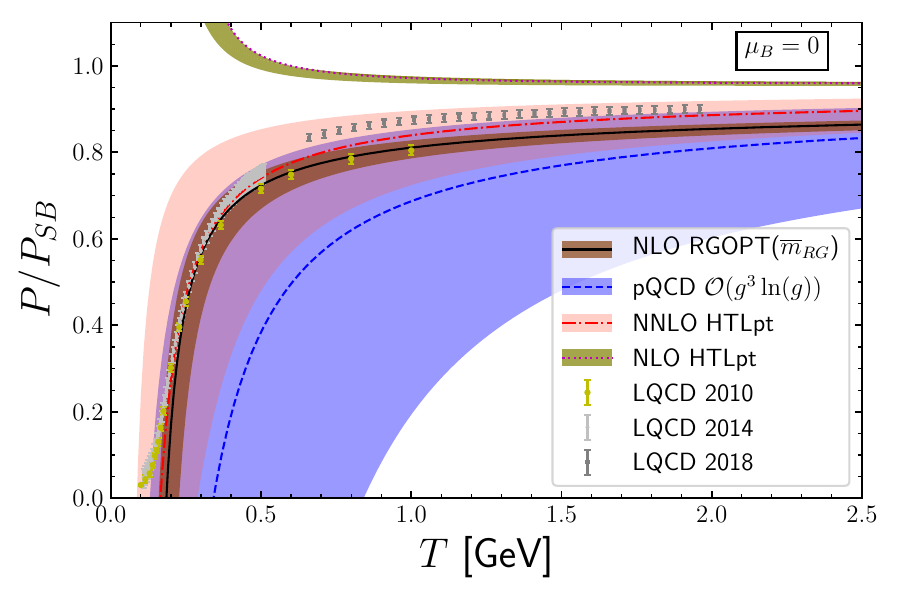,width=1.\linewidth,angle=0}}
\caption{ NLO RGOPT (RG prescription) plus NLO $P^{PT}_g$ pressure (brown band) 
compared to  ${\rm N}^3{\rm LO}\, g^3 \ln g$ pQCD (light blue band),
NLO HTLpt (light green band) and NNLO HTLpt (light red band), 
with scale dependence $\pi T \le M \le 4\pi T$, and to lattice data \cite{LQCD2010,LQCD2014,LQCD2018} at $\mu_B=0$.  }
\label{PRGprescg}
\end{figure}
Fig. \ref{PRGprescg} shows the RGOPT QCD pressure normalized by $P_{SB}\equiv P^q_{SB}+P^g_{SB}$ as a function of $T$ for $\mu_B=0$,
obtained with our best RG prescription, 
Eqs.(\ref{RGexact}),(\ref{B2sq0}). 
One notices that for $T \gtrsim 0.25 \, {\rm GeV}$ our results
 display a remarkable agreement with the LQCD data of \cite{LQCD2010} all the way up to $T=1\,{\rm GeV}$ 
(the highest value considered in those simulations), as well as a very good agreement 
with more recent LQCD data~\cite{LQCD2014} at intermediate $T$.
Furthermore the RGOPT results are drastically more stable than pQCD and HTLpt when $M$  is varied, 
as clearly indicated by the different band widths associated to the different approximations.
The NLO RGOPT results at central scale, $M=2\pi T$, observe a 
better agreement with LQCD data from \cite{LQCD2010} than NNLO HTLpt for 
$0.5\, {\rm GeV}\lesssim T \lesssim 1 \,{\rm GeV}$, 
while the latter lies closer to the higher $T$ data of \cite{LQCD2018} as compared to 
RGOPT, showing sizeable differences. A concomitant feature however is the 
visible tension between low~\cite{LQCD2010} and higher $T$~\cite{LQCD2018} LQCD data in their common 
range [NB we show only statistical uncertainties for LQCD, 
as given in publically available files\cite{LQCD2010,LQCD2014,LQCD2018}].   
In Fig.~\ref{PRGprescg} one also notices that HTLpt at NLO and pQCD  at ${\cal O}(g^3 \ln g$) are still far from
lattice data, moreover the NLO HTLpt stays close to SB limit at intermediate to low temperatures.
We further mention that while the RGOPT 
band width illustrated correctly reflects the {\em total} remnant scale uncertainty resulting from 
both the (RG resummed) $P_q(M)$ and (standard perturbative) NLO $P_g$, the {\em sole} $P_q(M)$ uncertainty would be 
roughly comparable~\cite{companion} to the lattice error bars for $ T \gtrsim 0.5 \, {\rm GeV}$. \\
Next for completeness in Fig.~\ref{Pmu0bandg} similar results are shown for
the other possible MOP prescription Eq.(\ref{mop}), compared to lattice data and highest order pQCD. 
As anticipated this prescription is somewhat
less efficient than the RG one, regarding the remnant scale
uncertainty as well as lattice data comparisons. Yet with respect to pQCD or to (NLO) HTLpt,
overall it also appears as a sharp improvement, keeping in mind our NLO approximation.
%%%%
%
%%%%%%%%%%%
\begin{figure}[h!]
    \centerline{ \epsfig{file=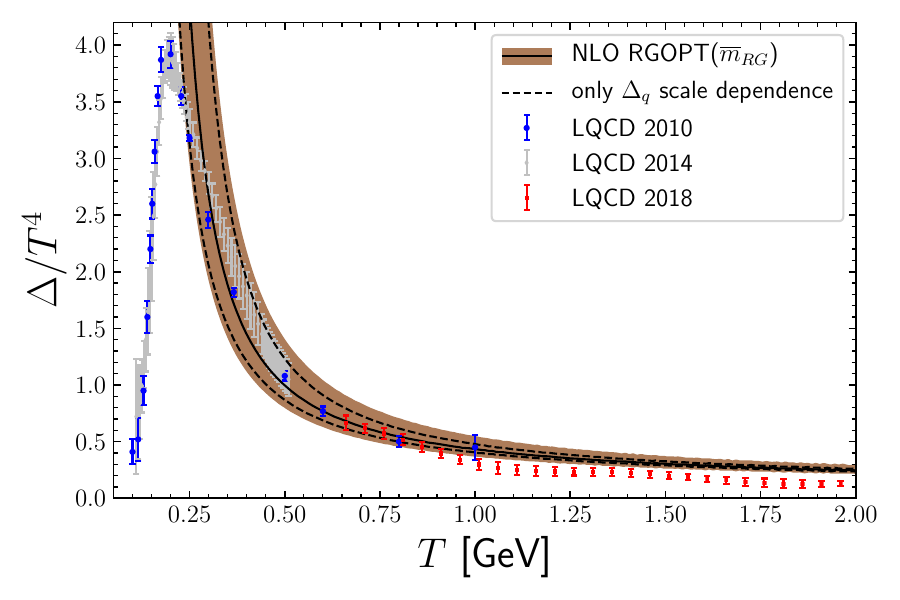,width=.9\linewidth,angle=0}}
\caption{ NLO RGOPT (RG prescription) trace anomaly $\Delta\equiv \varepsilon -3P$, 
including NLO $\Delta^{PT}_g$ (brown band), 
compared to lattice data \cite{LQCD2010,LQCD2014,LQCD2018}. 
The additional dashed lines illustrate the scale uncertainty originating solely
from RGOPT quark contributions within the full scale uncertainty  added by $\Delta_g^{PT}$ (brown) band.}
\label{TraceRG}
\end{figure}
%%%%%%%%%

Another physically interesting quantity is the interaction measure, $\Delta = {\cal E}- 3P$, [or at $\mu_B =0$ 
equivalently $\Delta =T^5\,\partial (P/T^4)/\partial T $, since ${\cal E} =-P + {\cal S} T$ with 
${\cal S}= \partial P/ \partial T$ representing the entropy density]. 
In Fig.~\ref {TraceRG} the NLO RGOPT interaction measure at $\mu_B=0$, 
obtained by a numerical derivative from our best RG prescription above, 
are compared to the available LQCD data \cite{LQCD2010,LQCD2014,LQCD2018}. 
At  temperatures $0.3 \,{\rm GeV} \lesssim T \lesssim 1 \,{\rm GeV} $ a 
very good agreement is observed.  However, similarly to pQCD and HTLpt, NLO RGOPT as applied here 
does not describe correctly the peak region near the pseudocritical $T_c$ temperature as exhibited by
lattice data. 

{\em Conclusions.} In this work we have compared RGOPT predictions regarding the QCD pressure with lattice data 
for the first time.  Our NLO predictions for the central scale, $M=2\pi T$, turned out to compare very well 
for temperatures starting at $T \simeq 0.25 \, {\rm GeV}$ which lies within the relatively strong coupling regime 
($\alpha_s \simeq 0.3$).  This agreement persists up to $T= 1\, {\rm GeV}$, 
the highest value for the LQCD data of \cite{LQCD2010}. Furthermore, comparing our NLO results with those from  
NNLO HTLpt one observes that the consistent RG invariance properties native to the RGOPT are  
drastically attenuating the remnant scale dependence issue. 
While the striking agreement with lattice data of \cite{LQCD2010} 
in Fig.~\ref{PRGprescg} may be partly numerically accidental, variants of our
prescription in Fig.~\ref{Pmu0bandg} still appear in very good
agreement given our essentially NLO-based construction, as compared to the state-of-the-art perturbative higher order QCD.
 The differences of our results with higher $1 \, {\rm GeV}\lesssim T \lesssim 2\, {\rm GeV}$ LQCD data\cite{LQCD2018} are 
however visible. Incidentally the LQCD results in \cite{LQCD2010} and in \cite{LQCD2018} appear to be in tension 
in their common range, while the trace anomaly shows more continuity,
a feature that may call for more investigations independently of our results.
Regarding the comparisons with $2+1$ flavor LQCD here illustrated, 
one may also keep in mind our presently not fully realistic
approximation of $N_f= 3$ degenerate flavors.
 We recall moreover that owing to present technical difficulties in readily 
applying our approach to the gluon sector, it is included as a purely perturbative NLO
contribution on top of the variationally resummed quark contributions.  
While this simple prescription appears to describe fairly well the moderate to high-$T$ regimes 
$T \gtrsim 0.25 \, {\rm GeV} \sim 1.5\, T_{pc}$, beyond NLO one could not avoid 
to face the well-known infrared divergences from gluon contributions, calling for appropriate resummations.
Finally we mention that the NLO RG-improved 
properties exhibited here extend without degradation to sizeable chemical potential 
values (for illustrations we refer to our companion paper \cite{companion}). 
The latter indicate the potential of our approach towards a more systematic exploration of both hot and dense QCD.

{\em Acknowledgments}:
 We thank P. Petreczky for bringing the results of Ref.~\cite{LQCD2018} to our attention.
M.B.P. and T.E.R. are partially supported by Conselho Nacional de Desenvolvimento Cient\'{\i}fico e Tecnol\'{o}gico 
(CNPq-Brazil) and by Coordena\c c\~{a}o  de Aperfei\c coamento de Pessoal  de  N\'{\i}vel Superior-(CAPES-Brazil)-Finance 
Code 001. This  work  was also  financed  in  part  by    INCT-FNA (Process No.  464898/2014-5). 

\end{document}